\documentclass[twocolumn,prX]{revtex4}
\usepackage{amsmath}
\usepackage[dvips]{color}
\usepackage{hyperref}
\usepackage{amssymb,amsmath,amsfonts,amsthm,epsf,mathrsfs}
\usepackage{graphicx}
\long\def\@makefntext#1{\parindent 0cm\noindent \hbox to
1em{\hss$^{\@thefnmark}$}#1}

\newcommand{\be}{\begin{equation}}
\newcommand{\ee}{\end{equation}}
\newcommand{\bea}{\begin{eqnarray}}
\newcommand{\eea}{\end{eqnarray}}
\newcommand{\pa}{\partial}

\newcommand{\ac}{\frac}

\newcommand{\lom}{\Omega}

\begin{document}
\title{Asymptotic Safety, Asymptotic Darkness, and the hoop conjecture in the extreme UV}
\author{Sayandeb Basu\\ 
\emph{Physics Department, University of the Pacific 3601 Pacific
Avenue,
Stockton, CA 95211}\\
 and\\
David Mattingly\\
\emph{Physics Department, University of New Hampshire, Durham, NH
03824}}

\begin{abstract}
Assuming the hoop conjecture in classical general relativity and quantum mechanics, any observer who attempts to perform an experiment in an arbitrarily small region will be stymied by the formation of a black hole within the spatial domain of the experiment. This behavior is often invoked in arguments for a fundamental minimum length. Extending a proof of the hoop conjecture for spherical symmetry to include higher curvature terms we investigate this minimum length argument when the gravitational couplings run with energy in the manner predicted by asymptotically safe gravity. We show that argument for the mandatory formation of a black hole within the domain of an experiment fails. Neither is there a proof that a black hole doesn't form.  Instead, whether or not an observer can perform measurements in arbitrarily small regions depends on the specific numerical values of the couplings near the UV fixed point. We further argue that when an experiment is localized on a scale much smaller than the Planck length, at least one enshrouding horizon must form outside the domain of the experiment. This implies that while an observer may still be able to perform local experiments, communicating any information out to infinity is prevented by a large horizon surrounding it, and thus compatibility with general relativity can still be restored in the infrared limit.
\end{abstract}

\maketitle

\section{Introduction}
The conjecture that the spectrum of particle states near the Planck
scale is dominated by black holes, so called ``asymptotic
darkness'', is an old one. Proof of such a conjecture requires a
particular theory of quantum gravity, however one can make an
argument with solely general relativity and quantum mechanics. The
argument relies on Thorne's hoop conjecture \cite{Thorne}: If an
amount of energy or matter $E$ is compressed into a region of
characteristic size $R$, a black hole forms if $R<R_S=2GE$. If $R$
is identified with the ``size'' of an experiment and taken to be the
De Broglie wavelength, which implies the scaling of size with energy 
$R\propto 1/E$, then a black hole forms for any experiment at
energies $E>E_{Planck}$. Choptuik and Pretorius \cite{Matt} have
recently verified numerically that in classical general relativity
the generic outcome of two particle scattering at small impact
parameters indeed is a black hole. This vindicates the insights
derived from the Aichelburg-Sexl result \cite{Peter,Aichelburg} and
other work on particle scattering ~\cite{giddings2, Fischler, Aharony,
Kaloper, Eardley}.

Asymptotic darkness has interesting consequences for the structure
of quantum field theory\cite{crane, Banks2}, as well as for the status of
gravity as an ordinary field theory in the UV.  In particular, there
is a relationship between black hole dominance and
non-renormalizability of gravity. The scaling of black hole entropy
with area implies that in asymptotically flat $d$-dimensions, the
high energy density of states in black holes grows with energy as
$e^{E^{\ac{d-2}{d-3}}}$. This scaling behavior
then implies that quantum gravity cannot be related to a local
conformal field theory in the UV, since in the latter case the
universal scaling of density of states goes as
$e^{E^{\ac{d-1}{d}}}$. The implication, as pointed out in ref.
\cite{shomer}, is that gravity cannot be an ordinary renormalizable
field theory, since a renormalizable field theory can be considered
to be a perturbation of a conformal field theory by relevant
operators, whose couplings at a given scale remain finite.

Asymptotic darkness also comes into play for some arguments
proposing a minimum observable length \cite{Calmet}.  If a black
hole necessarily forms as one shrinks an experiment below the Planck
length, then the Planck length serves as a minimum measurable
distance as one cannot extract any information from the black hole.
 The existence of a fundamental length scale is considered by many a
fairly model independent feature of quantum gravity \cite{Garay} and
could provide a fundamental cutoff to any quantum field theory.

In short, if asymptotic darkness is correct there are qualitatively
new features for gravity in the UV that are not present if we simply
think of gravity as a local quantum field theory of a spacetime
metric. What then happens in a scenario for quantum gravity where
gravity \textit{is} supposed to be a well-behaved local quantum
field theory in the UV as well as the IR?  Such a scenario is
envisioned by the asymptotic safety program, originally championed
by Weinberg \cite{Weinberg1, Weinberg2}. Asymptotic safety can be
described in the following heuristic way. Consider a truncated (for
technical reasons, see ~\cite{niedermaier2}) effective action for 4-d
metric gravity with cosmological constant coupled to matter
\begin{align}
I=\int d^4 x  \sqrt{-g}\big(- \Lambda(\mu) + \kappa_0(\mu) R+ \kappa_1(\mu)R^2+{}\nonumber\\
\kappa_2(\mu) R_{ab}R^{ab} + L_m \ldots\big) \label{eq:action}
\end{align}
containing a finite set of higher derivative terms with associated
coefficients, $\kappa_{1,2,3...}$ in addition to the usual
cosmological constant and Einstein-Hilbert terms.  $\mu$ is an
energy scale associated with an experiment and the couplings are
allowed to run with $\mu$. Since gravity couples to the matter
stress-energy tensor, the actual strength of the gravitational
coupling in a given experiment is (heuristically) totally controlled
by $\mu$.  It is therefore useful to consider the dimensionless
coupling $\bar{\kappa}_I$ between gravity and matter generated by
each term, $\bar{\kappa}_I=\mu^{d} / \kappa_I$, where
$d=[\kappa_I]$, i.e the mass dimension of the coefficient (as
opposed to the operator itself). So, for example, the
Einstein-Hilbert term generates a dimensionless coupling
$\bar{\kappa}_0=\mu^2/\kappa_0$, which is the familiar $G \mu^2$
interaction term. Higher curvature terms generate dimensionless
couplings with lower powers of $\mu$. Some of these dimensionless
couplings could blow up under the renormalization group flow of the
$\kappa_I$'s (with respect to the scale $\mu$). In particular, if
$\kappa_0$ did not run, $\bar{\kappa}_0$ diverges as $\mu
\rightarrow \infty$.  In asymptotic safety, the dimensionless
couplings remain finite and flow to a UV fixed point instead as $\mu
\rightarrow \infty$.  If this is true then metric gravity remains a
valid, and in principle predictive description, down to arbitrarily
short distances and high energies. While we do not know whether
quantum general relativity is asymptotically safe, quite a few
truncations offer evidence of a UV fixed point \cite{Reuter, Niedermaier, niedermaier2, Litim, benedetti}.

Since asymptotic safety requires that quantum gravity is a local field theory valid to arbitrarily high energies, while asymptotic darkness indicates that quantum gravity is not, there is an apparent tension between these two ideas.  In this work we investigate this problem by re-examining the hoop conjecture in the context of the minimum length argument using one of the results of
asymptotic safety - the running of coefficients such that
$\bar{\kappa}_I$ remains finite.   In particular we concern ourselves
with the necessity of one or more horizons forming for an arbitrarily localized experiment and, more importantly, where they form.  We then leverage a subtlety to the minimum length argument that allows us to perhaps reconcile asymptotic safety with asymptotic darkness for large black holes, as we now discuss.  

Obviously, since any theory of quantum gravity must reduce to general relativity, a black hole must form for an experiment localized to scales much smaller than the Planck length as the necessary energy would be much greater than the Planck energy.  The  resulting black hole would be large compared to the Planck length and so governed by ordinary general relativity.  This UV/IR connection has been pointed
out by Banks among others~\cite{Aharony, Banks2}. Formation of such
a hole would certainly preclude transmission of any information
generated by a localized experiment out to infinity, however it does
not a priori imply that an experiment cannot locally measure
arbitrarily small scales. Forbidding the measurement of arbitrarily
small scales requires the horizon to be \textit{inside} the region
of the experiment. Consider for the moment pure general relativity
and spherical symmetry, such that the resulting black hole is
Schwarzschild.  In this case any 2-sphere inside the black hole
necessarily contains an apparent horizon from which information
cannot escape and there is a minimum length.  However, for Planck
sized regions we are not in pure general relativity anymore - the gravitational constants may run significantly and higher curvature terms become relevant. Hence the small scale geometry may change significantly. If, for example, a pair of horizons formed outside a localized
experiment then there would be no trapped surface inside the region
of the experiment and therefore no minimum length. A highly localized experiment would still create a black hole as seen by an observer at infinity,
but that is irrelevant to arguing that the experiment can not
resolve distances below the Planck length.

For our simple thought experiment, consisting of a ball of uncharged matter
inside a sphere $\lom$ of proper radius $L$ with total energy $E
\propto 1/L$, the necessity of a horizon forming inside $\lom$
as $L \rightarrow 0$ vanishes for asymptotic safety. There is no
theorem that it \textit{doesn't} form, and in fact whether it does
or not depends on the size of higher curvature terms which are
currently unmeasured and not uniquely calculated in asymptotic safety, but the strong argument from general
relativity and quantum mechanics no longer holds.  Hence the minimum
length argument for localized experiments falls apart. We further
conjecture that for $L<<L_{Planck}$ an interior horizon forms
\textit{outside} of $\lom$ but with radius less than $L_{Planck}$. While this would not necessarily have
an effect on local experiments, it does mean that one cannot
transmit information from the local experiment to infinity, as we
noted above.  The $L<<L_{Planck}$ states also have a large outer
horizon that approaches the ordinary Schwarzschild horizon of
general relativity. This is similar to the modified Schwarzschild solution of Reuter and
Bonanno~\cite{ReuterBonanno} that possesses an interior and exterior
horizon.\footnote{We caution the reader that in our approach the ``inner horizon'' is not a true spacetime inner horizon in the sense of a Reissner-Nordstrom inner horizon. It is instead rather experiment dependent, in that the location of the inner horizon depends on the total energy of the spacetime and the specific total energy of an experimental probe.}  In this way, the extreme UV behavior remains possibly
compatible with both asymptotic safety and the results of general
relativity for large black holes where quantum gravity is not
important.  A local experiment may not measure a minimum length, but
we can't get that result back to our low energy, IR world that lives
outside the outer horizon.

\section{The hoop ``theorem'' for spherical symmetry}
Black hole dominance in the UV is a rather vague statement - there
are myriad ways one could try and form black holes.  However,
if the hoop conjecture is correct, pretty much any way we
concentrate enough matter into a small enough region should form
one.  Therefore we choose a specific behavior for gravity, our
spacetime topology, and the distribution of matter as follows:

\begin{enumerate}
  \item Gravity is asymptotically safe.
  \item The spacetime topology is such that we can foliate spacetime
  into spacelike slices $\Sigma_t$, where $t$ is a time function.
  \item A spherical ball $\lom$ of proper
  radius $L$ is filled with matter with density $\rho$  on a maximal spacelike slice $\Sigma_0$.
  \item The system is at least momentarily stationary.

\end{enumerate}

If an experiment of proper radius $L$ is to exhibit a minimum
length, then there must be at least an apparent horizon
\textit{inside} $\Omega$ as we shrink the size of the experiment.
Our question is then whether an apparent horizon must be present on
$\Sigma_0$ inside $\Omega$ in the limit as $L\rightarrow 0$.

We choose this particular setup for a number of reasons. First, the
hoop conjecture in its usual form rests on explicitly using the
Schwarzschild radius of a black hole. This poses several different
problems. First, the event horizon is a global object. Further, the
very existence of a sharply defined event horizon is questionable in
a theory of quantum gravity. This observation  has prompted much
work on reformulating black hole physics in terms local trapping
and/or dynamical horizons~\cite{Beetle, Ashtekar}. We therefore use
a more local reformulation of the conjecture in terms of trapped
surfaces \cite{bizon,flanagan}/apparent horizons \cite{Frolov,
Nambu, Chiba}. Spherical symmetry and the momentarily stationary
assumption are both used for simplicity. We note, however, that the
stationary assumption also intuitively matches the idea of scaling a
static system down to try to measure ever smaller distances in some
frame.

The proof of Bizon et. al.~\cite{bizon}, uses the Hamiltonian
constraint from classical general relativity and constitutes a proof
of the minimum length argument under certain conditions. We simply
adjust the proof to include higher curvature terms and the running
of coefficients. One might argue that the classical
Hamiltonian constraint shouldn't be applied for questions about
quantum gravity.  Generically, there could be quantum
corrections or other modifications to the Hamiltonian constraint at
the Planck scale that are not captured by this simple semiclassical approach. However, we take the view that since we are asking how asymptotic safety can modify the standard minimum length argument, which is based on just such a semiclassical picture, the right thing to attempt first is to consider a specific prediction of asymptotic safety on a specific,
mathematically precise formulation of the argument.  A more problematic concern is specific to asymptotic safety itself, which is defined perturbatively.  We use the results of a fully quantum mechanical but perturbative framework (where there is a well-defined S-matrix) to gain insight into a non-perturbative question, and there is no guarantee that non-perturbative effects can or should be neglected.  However, this is again exactly what is done in the usual minimum length argument, where one treats an effective field theory (general relativity) that is only consistent in the low-energy perturbative regime classically even in the extreme UV.  Hence for now we proceed in the same spirit and turn to the proof itself.

Consider a foliation of spacetime with Cauchy hypersurfaces denoted
by $\Sigma_t$. Let $n^a$ be the normal to the leaves of the
foliation. Choose $t=0$ and let $T^{ab}_{M}$ denote the matter
stress tensor and $\rho=T^{ab}_{M} n_a n_b$ the energy density of a
ball of matter in $\Sigma_0$.  We assume the matter is
instantaneously at rest inside a volume $\lom$ of \emph{proper}
radius $L$ bounded by a 2-surface $\pa \lom$.  The total proper
energy of the matter, $E$, is defined by the following integral
over proper volume $\lom$,
\begin{align}
E=\int_{\lom} T^{ab}_{M} n_a n_b=\int_{\lom} \rho dv.
\end{align}

$\pa \lom$ has a unit spacelike normal $r^a$. The spatial metric
$q_{ab}$ on $\Sigma_0$ and the extrinsic curvature $K^{ab}$
constitute canonical data. Whatever the particular truncation used,
any higher curvature terms in eq. (\ref{eq:action}) contribute to the
Hamiltonian constraint equation. We package the contribution of
those terms into a single quantity $H_\kappa$. Then the Hamiltonian
constraint equation for the pair $(q_{ab}, K^{ab})$ is given by
\begin{align}
^3R+K^2 -K_{ab} K^{ab}&= {\kappa_{0}}^{-1}\left( H_\kappa + \frac{1} {2} \rho\right).\nonumber\\
\label{constraints}
\end{align}
Note that the coefficients of the higher curvature terms as they
appear in the action eq. (\ref{eq:action}) are implicitly absorbed in
$H_\kappa$ in eq. (\ref{constraints}).  We do not consider the
contribution of the cosmological constant term to the Hamiltonian
constraint. The implication of setting $\Lambda=0$ is discussed
below when we review the renormalization group running of Newton's constant.

We first simplify using our assumptions.  For a maximal slice that
has everything momentarily stationary, the extrinsic curvature
$K_{ab}$ (and any other single time derivative) vanishes.  For
spherical symmetry, a gauge can be chosen such that the metric
$q_{ab}$ is conformally flat, $q_{ab}=\varphi^4(r) \delta_{ab}$.
This in turn implies that the 3-d Ricci scalar is given by $^3R=-8
\varphi^{-5} \nabla^2 \varphi$. For outgoing null rays emanating
from any 2-sphere inside $\Omega$ at radius $r$, the criterion that
the 2-sphere is not a trapped surface is a condition on the null
expansion
\begin{align}
\theta=r^a_{~;~a}> 0. \label{expansion}
\end{align}
Conversely, this must be less than or equal to zero if the 2-sphere
is a trapped surface. With $r^a_{~ ;a}=(r^2 \varphi^6)^{-1}
\ac{d(r^2 \varphi^4)}{dr}$, the condition eq. (\ref{expansion}) that
the 2-sphere is not trapped is simply
\begin{align}
\theta_r=\pa_r \left(r\varphi^2\right)  > 0~, \label{expansion2}
\end{align}
where the prime denotes an r-derivative.Finally, the proper radius
of $\pa \lom$ is $L=\int\varphi^2 dr$.

Now, fix the 2-surface $\pa \lom$ to have coordinate radius $r_0$
(and hence proper radius $\int_{0}^{r_{0}} \varphi^2 dr$).    Note
that if the energy density due to matter and that of the higher
curvature terms \cite{strominger, deser, deser2} is non-negative,
then from the Hamiltonian constraint
\begin{align}
^3R=-8 \varphi^{-5} \nabla^2 \varphi={\kappa_{0}}^{-1}\left(
H_\kappa + \frac{1} {2} \rho\right)
\end{align}
we have that
\begin{align}
\nabla^2 \varphi \leq 0 \label{laplace}
\end{align}
which in turn  implies that
\begin{align}
\ac{d \varphi}{dr}\leq 0. \label{dlessthanzero}
\end{align}
As Bizon et al. argue,  this condition is crucial in how the
integrated constraints lead to eq. (\ref{eq:Hoop2}) (see below).
This is rather obvious physically, as a negative energy density
would certainly allow one to avoid creating a black hole.
Integrating the Hamiltonian constraint over $\Omega$ we have,
\begin{align}
&\int_\Omega dv  (-8 \varphi^{-5} \nabla^2 \varphi) \\
\nonumber &= -\int_0^{r_0} 4 \pi \varphi^6 r^2 dr (8 \varphi^{-5}
\nabla^2 \varphi)\\ \nonumber &= (2\kappa_0)^{-1} \int_\Omega dv
\left(\rho + 2H_\kappa \right). \label{intconstr}
\end{align}
The right hand integral is $1/2\kappa_0$ times the total proper
energy $E$ of the matter contained within $\Omega$ plus the contribution
from the higher curvature terms. In spherical coordinates we then
have
\begin{align}
-32 \pi \int_0^{r_0} dr \varphi \partial_r(r^2 \partial_r
\varphi)=\frac {1} {2\kappa_0} E+\kappa_0^{-1} \int_\Omega dv
H_\kappa.
\end{align}
We can rewrite the left hand side of this equation as
\begin{eqnarray}
16 \pi \int_0^{r_0} dr \big( \varphi^2 - \partial_r (r \partial_r(r
\varphi^2))  \nonumber\\
+r \varphi \partial_r \varphi + \varphi^{-1} r (\partial_r \varphi)
\partial_r(r \varphi^2) \big)
\end{eqnarray}
The first term of the integral is simply $16 \pi L$, where $L$ is
the proper radius of $\Omega$.   We hence have
\begin{eqnarray} \label{eq:Hoop1}
16 \pi L - 16 \pi r \partial_r(r \varphi ^2)|_{d\Omega} \\
\nonumber + 16 \pi \int dr \left( r \varphi \partial_r \varphi +
\varphi^{-1}
r (\partial_r \varphi) \partial_r(r \varphi^2) \right) \\
\nonumber =\kappa_0^{-1} \left( \frac {E} {2} + \int_\Omega dv
H_\kappa \right) .
\end{eqnarray}
If there are no trapped surfaces in $\Omega$ and $\pa \Omega$ is not
a trapped surface then by eqs. (\ref{expansion2}) and
(\ref{dlessthanzero}) every term except the first on the left hand
side of (\ref{eq:Hoop1}) is negative.  Therefore if there are no
trapped surfaces in $\Omega$ we have the condition that
\begin{eqnarray} \label{eq:Hoop2}
16 \pi L > \kappa_0^{-1} \left( \frac {E} {2} + \int_\Omega dv
H_\kappa \right) .
\end{eqnarray}
The converse is also true - if this inequality is violated there is
a trapped surface in $\Omega$.  This then is the revised hoop
condition that must be met by the mass and the integrated density of
higher curvature terms.

In the case of pure general relativity, i.e. with $H_\kappa=0$ and
no running of couplings, eq. (\ref{eq:Hoop2}) in conjunction with
quantum mechanics implies that a trapped surface will inevitably
form as the size of an experiment is shrunk.  As we reduce the
proper distance we are trying to resolve, $E$ must rise according to
the uncertainty principle.  If we trap an object in a region of
proper distance $L$, $E$ scales as $L^{-1}$.  In the limit as $L
\rightarrow 0$ the LHS of eq. (\ref{eq:Hoop2}) goes to zero while
$E$ diverges.  Hence at some point the inequality cannot be
satisfied and a trapped surface must form. This occurs at energy
$E=32 \pi \kappa_0 L \approx 32 \pi \kappa_0 E^{-1}$, i.e. at energy
$E \approx \sqrt{\kappa_0}=E_{Planck}$. At energies greater than
$E_{Planck}$ the situation only gets worse, as a horizon is still
present inside $\Omega$.

\section{Implications of asymptotic safety}
In the asymptotic safety scenario of quantum gravity, Newton's
constant, $G=(16 \pi \kappa_0)^{-1}$ is itself a running coupling
(eq.~\ref{eq:action}). We first identify the energy scale $\mu$ of
the running with $E$, the total proper energy in our ball of matter.
This is of course not exactly true.  Since we have a localized
experiment there is no one particular energy we can assign to it (in
contrast to a scattering scenario, for example).  However, the total
energy $E$ still is acceptable as a rough estimate of the energy
scale we are probing spacetime at, especially since we are relating
$E$ to the proper length $L$ via $E \sim 1/L$, i.e. just interested in the scaling behavior.

A number of recent works in the literature \cite{benedetti} have
confirmed the existence of a non-gaussian fixed point with various
higher curvature truncations and matter contribution included in the
total effective action. Integrating the flow equation is an involved
process, but the upshot that we will exploit in the present article
is that in the extreme UV, the running of Newton's constant is
dominated by the fixed point behavior. This is given by \be
\ac{1}{16 \pi \kappa_0(E)}=G(E)\simeq \ac{g_{*}}{E^2}
\label{running} \ee where $g_{*}$ is the value of the dimensionless
Newton's constant at the non-Gaussian fixed point, and we work in the
regime where the matter proper energy exceeds the fiducial Planck
energy set by the infrared value of Newton's constant.  Thus in the extreme UV as
$E^2$ increases, Newton's constant as a running coupling gets
progressively smaller. We also note in our discussion here and what follows, the energy scale $E$ is to be compared with fiducial Planck scale $E_{pl}$ (used below) which is defined with respect to the infrared value of Newton's constant $G_0=\ac{1}{8 \pi E_{pl}^2}$.

\subsection{The cosmological constant}
Before we examine the effect of this behavior on the formation of
local trapped surfaces, we would like to justify setting
$\Lambda=0$. As discussed in existing literature
\cite{niedermaier2}, when we consider pure gravity, with say just
the Einstein-Hilbert term, Newton's constant is an inessential
coupling and it can be changed by an overall field redefinition.
Such redefinitions will leave the action invariant. Thus in the pure
gravity sector, the running of Newton's constant is usually
understood with reference to the running of some other fiducial
operator in the action. The most commonly adopted choice for this
reference operator is the cosmological constant term. In detail, the
RG flow equation for the dimensionless Newton's constant
\begin{align}
E\ac{\pa g}{\pa E}=(2+\eta)g \label{beta}
\end{align}
contains an anomalous dimension term $\eta(\Lambda)$. For the
Einstein-Hilbert truncation this is given by
\begin{align}
\eta^{EH}=\ac{g B_{1} (\Lambda)}{1-g B_2 (\Lambda)}
\label{anomalous}
\end{align}
Here $B_1(\Lambda)$ (respectively $B_2(\Lambda)$) are cosmological
constant dependent functions which additionally  also depend in a
rather involved way on the details of how the infrared cut-off for
the Exact Renomalization Group Equations (ERGE) is chosen. The
anomalous dimension can be evaluated at zero arguments for these
functions, $B_1(0)$ and $B_2 (0)$, effectively restricting to the
case of zero cosmological constant.  In the case of the
Einstein-Hilbert truncation the so-called optimized cut-off due to
Litim \cite{Litim} for instance leads to
\begin{align}
\eta^{EH}=\ac{12g}{2g-2(\Lambda-\ac{1}{2})^2}, \label{anomalous2}
\end{align}
 hence setting $\Lambda=0$ in
eq. (\ref{anomalous}) amounts to having
\begin{align}
\beta_{g}=\ac{(1-16g)2g}{1-4g}\label{runwithoutcosmo}
\end{align}
The behavior of Newton's constant gleaned from  eq.
(\ref{runwithoutcosmo}) near the non-Gaussian fixed point
$g_{*}=\ac{1}{16}$, for large $E$, is $G\simeq \ac{g_{*}}{E^2}$. Let
us also note in passing that when higher curvature truncations are
included in the effective action, the effect of these terms show up
again in the anomalous dimension, now written as $\eta(\Lambda,
H_\kappa)$. The fixed point behavior that this leads to has been
discussed by Benedetti et al. \cite{benedetti}.

\subsection{Running Newton's constant and the formation of trapped
surfaces} We now include the running of $\kappa_0$, eq.
(\ref{running}) into eq.(\ref{eq:Hoop2}). Momentarily restricting to
the Einstein-Hilbert + matter truncation only we find the condition
for no trapped surfaces inside $\lom$ to be
\begin{align}
 L> \ac{g_{*}}{2E^2}~ E
\label{hoopwithouth}
\end{align}
Now using the uncertainty relation, we replace $L\sim1/E$. The
factors involving the proper energy $E$ drop out from both sides of
the inequality eq. (\ref{hoopwithouth}) and the (necessary)
condition for \emph{no} trapped surfaces to lie in $\lom$ takes the
form
\begin{align}
g_{*}<2 \label{constronfixedpoint}
\end{align}
This condition shows that up to $O(1)$ numbers arising from the order of magnitude estimates/scaling arguments we use in our derivation, there is \textit{some} value of coefficients for which the minimum length argument falls apart, in contrast to ordinary general relativity.  Interestingly, $g_*$ in different truncations has been found to be quite close to $2$.  For example, the most recent results of Benedetti et al. \cite{benedetti}, who consider both a minimally coupled scalar field with the Einstein-Hilbert action and a higher curvature truncation involving the Ricci and Weyl scalars, are
\begin{align}
g_{*}&=0.860 ~~~(\textrm{For Einstein-Hilbert truncation}){}\nonumber\\
g_{*}&= 2.279 ~~~(\textrm{For}~R^2+C^2~\textrm{truncation})
\end{align}
In general the inclusion of higher curvature invariants in the effective action, constructed out of Ricci, Riemann and Weyl tensors, pushes the high energy fixed point towards values higher than their Einstein-Hilbert counterparts. 

Consider now the impact of higher curvature terms with running couplings on our argument.
Returning to eq. (\ref{eq:Hoop2}) we find that no trapped surfaces
form inside $\lom$ if
\begin{align}
L ~> \ac{g_{*}}{E^2}\left(\ac{{E}}{2}+\int dv H_\kappa \right)
\label{eq:Hoop2a}
\end{align}
With $L$ replaced by $E^{-1}$ eq. ~(\ref{eq:Hoop2a}) becomes
 \label{eq:Hoop3} \begin{align}
E^{-1}>\ac{g_{*}}{E^2}\left(\ac{E}{2}+\int H_\kappa dv\right)
\end{align}
We now observe that the integral involving the higher curvature
terms is dimensionally an energy and so the integral must be a
combination of $E$ multiplied by the dimensionless coefficients
$\bar{\kappa}_I$. We therefore have,
\begin{align}
\label{eq:Hoop4}
E^{-1}>g_{*}\left((2E)^{-1} + E^{-1}
\bar{I}(\bar{\kappa}_{I})\right)
\end{align} where
$\bar{I}(\bar{\kappa}_I)$ is a dimensionless number whose details
depend on both the unknown $\kappa_I$ and the precise solution for the metric in the presence of the
higher curvature terms with our given stress tensor. It suffices for
us that it is a linear function of $\bar{\kappa}_I$. In asymptotic
safety the $\bar{\kappa}_I$ coefficients remain finite as $E
\rightarrow \infty$, and hence $I$ is also finite.  Therefore as
$E\rightarrow \infty$ there is no proof that a horizon is formed, as
each side scales the same way with $E$, and so the argument for a minimum length no longer necessarily holds (at least in this formulation).

Of course, in the above approach, there is no proof that a horizon
is \textit{not} formed either.  Since $E$ drops out in asymptotic
safety at high energies, whether or not a horizon forms inside $\Omega$ depends on
both the particulars of the value of $g_*$ at the non-Gaussian fixed point
(and therefore on the truncation) and the actual size of the
$\bar{\kappa}_I$ coefficients.  Since both eq. (\ref{hoopwithouth}) and eq. (\ref{eq:Hoop4}) are obtained in the extreme UV limit in which we invoke the largely scheme independent \cite{robert2} fixed point dominated running eq. (\ref{running}), our conclusions (the bound on the fixed point $g_{*}<2$, modulo $O(1)$ numbers discussed above) are at least essentially insensitive to myriad IR-cutoff schemes in vogue and the precise size of the higher curvature terms at some fiducial infrared scale.

\section{General relativity and the infrared limit}
While the question of whether or not a horizon forms inside $\Omega$
for Planck sized experiments becomes muddled with asymptotic safety,
in the limit of energy much greater than $E_{Pl}$ there still must
be a trapped surface at large radius, as this corresponds to a large black hole with low horizon
curvature (hence governed by general relativity).  To reach such a spacetime picture in the semiclassical limit, one considers quantum states with low enough energy density that gravitational backreaction can be neglected propagating in a spacetime generated by some separate matteer distribution.  For our super-Planckian experiments, we have no such luxury as we specifically are exploring a change in the gravitational action, the running of coefficients, via the energy scale of the experiment.  However, we can analyze the ``horizon'' structure generated by a matter distribution within our framework by considering a background distribution of matter with total energy $E_B$ in $\Omega$ and ask when an experiment at a length scale $L_E$ (or equivalent energy scale $E_E$) encounters a trapped surface.  

If we wish to run our experiment all the way back to the infrared, we can no longer consider just the UV fixed point behavior for $G$.  We therefore will take a specific form for $G(E)$ \cite{ReuterBonanno, hewett},

\begin{equation}
\label{eq:GofE}
G(E)=\frac{g_*}{E^2 + g_* M_{Pl}^2}
\end{equation}

and return to equation (\ref{eq:Hoop2}), but modify the
scenario.  We define

\begin{equation}
F(E_E) = L_E - G(E_E) \left(\ac{{E_E + E_B}}{2}+\int dv H_\kappa \right)
\end{equation}
$F(E)>0$  is the condition for no trapped surface in the region.  As $E_E \rightarrow \infty$ it dominates this expression, so we return to our original formulation of the hoop conjecture for a single experiment.  We assume that $g_*$ and the $\kappa_I$ are such that there is no trapped surface within $\Omega$ as $E_E \rightarrow \infty$.  Given this, as we $E_E$ towards the IR eventually $F(E_E)$ may cross zero.  This signifies that at some radius $L_E$ we have crossed into a region that contains trapped surfaces.  The inner boundary of this region can be thought of as an ``inner horizon'', similar to the Reissner-Nordstrom inner horizon, although we caution that it is defined only as a function of the experiment energy $E_E$ and so is not a true spacetime quantity.  Similarly, at very small $E_E$, $F(E_E)$ can cross zero again, signaling the outer boundary of our trapped surface region.  This outer boundary surface corresponds to the black hole event horizon in the infrared.

We now solve for the location of these two boundary surfaces.  The presence of a horizon is when 

\begin{equation} \label{eq:FE0}
L_E - G(E_E) \left(\ac{{E_E + E_B}}{2}+\int dv H_\kappa \right)=0
\end{equation}

$E_E$ and $L_E$ are related via $E_E = 1/L_E$, but $E_B$ is to be considered fixed.  The contribution of the higher curvature terms can be re-expressed as $I_E(\kappa_I) E_E$ and $I_B(\kappa_I) E_B$, where each $I$ is again a dimensionless number dependent on the $\kappa_I$'s. If we define $J_E=1/2 + I_E$ and $J_B=1/2 + I_B$ we can rewrite (\ref{eq:FE0}) in terms of the $J$'s and $L_E$ as

\begin{equation}
L_E-\frac {g_* L_E^2} {1+g_*L_E^2 M_{Pl}^2} \left(\frac {J_E} {L_E} + J_B E_B\right)=0.
\end{equation} 

Solving for $L_E$ yields

\begin{equation}
L_E=\frac{g_* E_B J_B \pm \sqrt{g_*^2 E_B^2 J_B^2 - 4 g_* M_{Pl}^2(1-g_* J_E)}} {2 g_* M_{Pl}^2}.
\end{equation}

We now consider the ``large black hole limit'' in the infrared by assuming $E_B=NM_{Pl}$ with $N>>1$, i.e. there's a lot of mass in our background spacetime. In this limit we have

\begin{eqnarray}
L_{E+}= \frac {E_B J_B} {M_{Pl}^2}\\
L_{E-}=\frac{1-g_* J_E}{g_*E_BJ_B}.
\end{eqnarray}
$L_{E+}$ is $R \approx GE_B$, i.e. it corresponds to the Schwarzschild radius for a large black hole.   $L_{E-}$ is the location of the inner horizon.   We can get a more intuitive picture by setting all the higher curvature terms equal to zero.  In this case $J_E=J_B=1/2$ and the inner horizon location becomes $L_{E-}=(2-g_*)/(g_* E_B)$.  In the previous section, for just one experiment, our limit was roughly $g_*<2$ to avoid the minimum length argument.  Here we have the additional background energy $E_B$ in addition to $E_E$, so we must further decrease $g_*$ to avoid a horizon forming in the domain of an experiment. In other words, if we chose $E_E=E_B$ for example, then there is twice the energy density as we had in the previous section in a region of size $L_E=L_B$ so $g_*$ would need to be less than one to prevent formation of a horizon.  (Note that this experiment dependence of the limit on $g_*$ is symptomatic of this approach - whether and where horizons are formed depends strongly on how you plan to probe the spacetime as that influences the running of couplings.) If we consider $g_*<1$, such that this bound is satisfied, then we see that the inner horizon is reached when $L_E \approx 2/(g_* E_B)$.  We therefore have the following picture for $g_*<1$.  We have a background energy density $E_B>>M_{Pl}$ in a region of radius $L_B=1/E_B$.  We probe the system at length scales smaller than $L_B$ and we see no trapped surface.  As we approach the length scale $(2-g_*)L_B/g_*$ we see a trapped surface form inside our experiment. This behavior persists until we reach the large radius $GE_B$, at which point we return to semiclassical physics outside an ordinary Schwarzschild horizon.

\section{Conclusions}
There's a rather obvious tension between black hole dominance and the asymptotic safety scenario of quantum gravity. The black hole dominance arguments indicate that the very high spectrum is dominated by states whose entropy density, via the Bekenstein-Hawking formula, does \emph{not} scale as an extensive quantity with volume, which is a characteristic feature of any local conformal field theory. Since the high energy spectrum of renormalizable field theories are that of a conformal field theory, the implication is that gravity must be an exception to this rule. On the other hand, the asymptotic safety approach claims that precisely at those high energies where black holes dominate in the usual picture, a sensible local field theory based on the metric exists, and provides a reliable fundamental framework for quantum gravity. 

Motivated by this, we re-examined the hoop conjecture in light of asymptotic safety. Adopting an approach relatively independent of any specific truncation of the effective action, we explored how in particular one key feature of asymptotic safety, the running of couplings, affects standard arguments for a minimum length based on the hoop conjecture when the size of an experiment is shrunk below a critical length. For a gedanken experiment consisting of a ball of matter inside a volume $\lom$ of proper size $L$ and energy $E\sim \ac{1}{L}$, the fixed point behavior of the running couplings conspire such that the necessity of forming a trapped surface inside $\lom$ as we shrink the size of an experiment vanishes.  Hence the standard minimum length argument fails to necessarily be true in this case.  However there is no proof in this approach that a trapped surface does not form inside $\Omega$ - whether it does or not depends on specifics of quantum gravity that we do not know.
 
The qualitative features of the fixed point dominated running and indeed the existence of a non-Gaussian fixed point has been argued by Percacci \cite{robert2} to be a scheme independent feature of the the Wilsonian ERGE. He further notes that the position of the fixed point has a weak dependence on parameters in the choice of the IR cut-off function and so the values for the fixed points are thus believed to be quite robust. However, since our result depends on the fixed points in such a sensitive way, it is perhaps of general interest to extend the fixed point calculations to more realistic choices of matter Lagrangian and compare to a more detailed gedanken experiment.

Finally compatibility with general relativity can still be maintained even in our scenario. First, even if one can localize an experiment without formation of a trapped surface inside the region of the experiment, that is no guarantee that such microscopic information can be transmitted out to an observer at spatial infinity with physics dictated by general relativity.
In fact, we found that no information about the local physics can be communicated to an observer at infinity, even if a local experiment inside that horizon can in principle measure arbitrarily small length scales.  If we interpret asymtotic darkness as a statement about limitations on information transfer from a local experiment to an asymptotic observer owing to horizon formation, then our observations show that it may be possible to satisfy asymptotic darkness within the asymptotic safety scenario.  Furthermore, for spacetimes with a background total energy $E\gg E_{pl}$, where the physics of general relativity dictate that there must exist a large horizon at the classical Schwarzschild radius, we argue that as one probes at lower and lower energy scales (from the UV to the IR) one eventually encounters a trapped surface.  The resulting horizon structure consists of an experimentally dependent ``inner horizon'' and the classical Schwarzschild horizon in the extreme infrared.

\acknowledgements We thank Jishnu Battacharyya and Keisuke Juge for comments and stimulating discussions.

\end{document}